\documentclass[11pt, a4paper]{article}

\usepackage[margin=1in]{geometry}
\usepackage[T1]{fontenc}
\usepackage{graphicx}
\usepackage{amsmath}
\usepackage{booktabs}
\usepackage{multirow} 
\usepackage[table]{xcolor}
\usepackage{times} 
\usepackage{pbox}
\usepackage{amssymb}  
\usepackage{amsfonts} 
\usepackage{makecell}
\usepackage{caption} 
\usepackage{hyperref} 
\usepackage{authblk}
\usepackage{rotating}
\usepackage{array}
\usepackage{float}
\usepackage{enumitem}
\usepackage{multicol}

\title{\textbf{MammoClean: Toward Reproducible and Bias-Aware AI in Mammography through Dataset Harmonization}}

\author[1]{Yalda Zafari} 
\author[2]{Hongyi Pan}
\author[2]{Gorkem Durak}
\author[2]{Ulas Bagci}
\author[3,4]{Essam A. Rashed}
\author[1]{Mohamed Mabrok \thanks{Corresponding Author (\texttt{m.a.mabrok@gmail.com})}}

\affil[1]{Department of Mathematics and Statistics, Qatar University, Doha, Qatar}
\affil[2]{Department of Radiology, Northwestern University, Chicago, IL, United States}
\affil[3]{Graduate School of Information Science, University of Hyogo, Kobe 650-0047, Japan}
\affil[4]{Advanced Medical Engineering Research Institute, University of Hyogo, Himeji 670-0836, Japan}

\date{}
 
\begin{document}
\maketitle
\begin{abstract}
The development of clinically reliable artificial intelligence (AI) systems for mammography is hindered by profound heterogeneity in data quality, metadata standards, and population distributions across public datasets. This heterogeneity introduces dataset-specific biases that severely compromise the generalizability of the model, a fundamental barrier to clinical deployment. We present \textit{MammoClean}, a public framework for standardization and bias quantification in mammography datasets. \textit{MammoClean} standardizes case selection, image processing (including laterality and intensity correction), and unifies metadata into a consistent multi-view structure. We provide a comprehensive review of breast anatomy, imaging characteristics, and public mammography datasets to systematically identify key sources of bias. Applying \textit{MammoClean} to three heterogeneous datasets (CBIS-DDSM, TOMPEI-CMMD, VinDr-Mammo), we quantify substantial distributional shifts in breast density and abnormality prevalence. Critically, we demonstrate the direct impact of data corruption: AI models trained on corrupted datasets exhibit significant performance degradation compared to their curated counterparts. By using \textit{MammoClean} to identify and mitigate bias sources, researchers can construct unified multi-dataset training corpora that enable development of robust models with superior cross-domain generalization. \textit{MammoClean} provides an essential, reproducible pipeline for bias-aware AI development in mammography, facilitating fairer comparisons and advancing the creation of safe, effective systems that perform equitably across diverse patient populations and clinical settings. The open-source code is publicly available from: \href{https://github.com/Minds-R-Lab/MammoClean}{https://github.com/Minds-R-Lab/MammoClean}.

\end{abstract}

\section{Introduction}
\begin{figure}
    \centering
    \includegraphics[width=1\linewidth]{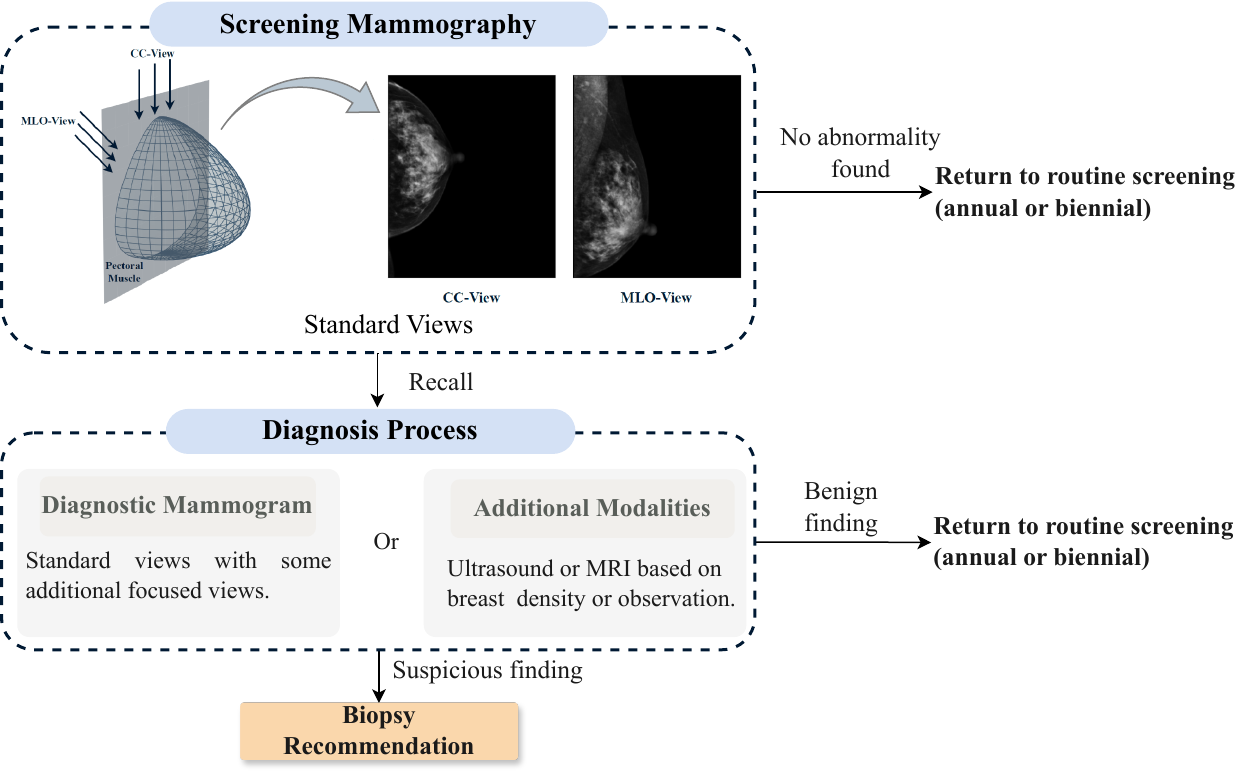}
    \caption{The role of mammography imaging in breast cancer screening and diagnosis.}
    \label{fig:mammography-role}
\end{figure}

Breast cancer is the most common cancer among women and a leading cause of cancer deaths \cite{wilkinson2022understanding}.  Regular screening with mammography is the most effective tool to detect breast cancer in early stages and prevent breast cancer-related deaths \cite{ ren2022global}. Therefore, mammography is the primary imaging tool for breast cancer screening and diagnosis, and it is also commonly used in follow-up \cite{duffy2002impact, broeders2012impact} (see Figure \ref{fig:mammography-role}). It uses low-dose X-rays to visualize breast tissue and underlying abnormalities, and the features of the resulting images vary depending on the imaging technique and the specific view captured. Among the available technologies, Digital Mammography (DM) or Full-Field Digital Mammography (FFDM) are the most widely used for producing two-dimensional (2D) mammographic images \cite{fiorica2016breast}. FFDM captures the entire breast in high-resolution grayscale images, providing faster interpretation compared to traditional Screen-Film Mammography (SFM) \cite{berns2006digital}. As a result, FFDM has largely replaced SFM in clinical practice.

The main limitation of FFDM lies in projecting the three-dimensional (3D) breast into 2D images, which inevitably causes information loss and tissue overlap. This is one of the main limitations of mammography and can cause early-stage cancers to be obscured behind dense and heterogeneous normal breast tissue. This challenge can be partially mitigated by acquiring and interpreting multiple views of the breast, which improves visualization of the 3D structure. Digital Breast Tomosynthesis (DBT), also known as 3D mammography, directly addresses this issue by capturing multiple angled images and reconstructing them into high-resolution 3D views \cite{chong2019digital}. DBT significantly improves cancer detection and reduces false positives; however, interpretation of DBT images generally requires more time compared to FFDM. Contrast-enhanced mammography (CEM) is another imaging method that uses a contrast agent to increase the visibility of abnormalities \cite{jochelson2021contrast}. By having a low-energy image similar to the standard mammogram and a high-energy image with contrast, the radiologist can analyze a more detailed view and can better detect and characterize abnormalities.

One of the most comprehensive approaches to mammography image analysis is the multi-view strategy. For each breast, two standard projections are acquired: the Cranio-Caudal (CC) view, captured from above the breast, and the Medio-Lateral Oblique (MLO) view, obtained from the side at an angle. Radiologists typically employ two main strategies, contra-lateral comparison and ipsi-lateral correlation, to identify corresponding findings across both views and to assess differences between the same views of the left and right breasts. This approach enables more reliable confirmation of suspicious findings and represents the most thorough method for interpreting a complete mammographic study.

The time-consuming interpretation time of mammograms, combined with the low prevalence of cancer in mammography images, makes deep learning models, an important subset of Artificial Intelligence (AI), a promising approach for automating parts of the clinical workflow and reducing the burden on healthcare professionals. Several studies have demonstrated encouraging outcomes from applying AI, and particularly deep learning, to mammography analysis. Reported results include cancer detection rates comparable to double reading without an increase in recalls \cite{laang2023artificial}, an improvement of up to 4\% in detection when replacing one radiologist with AI in double-reading setups \cite{dembrower2023artificial}, and reduced recall rates for low-risk cases \cite{park2025multi}. Several studies have developed and evaluated deep learning models for mammography analysis \cite{isosalo2023independent, manigrasso2025mammography, lin2025artificial, jeny2025hybrid, dai2026interpretable, zafari2025hybrid}, leveraging publicly available or private datasets.

A critical requirement for the development of AI-based models in mammography image analysis is the availability of standardized and harmonized data, which enhances both the training process and overall model performance. However, most datasets are collected, annotated, and formatted under heterogeneous protocols, resulting in inconsistencies in image quality, labeling standards, and metadata representation. Such variability creates challenges for reproducibility, interoperability, and ultimately the generalizability of AI models across diverse clinical environments. Well-processed and harmonized datasets not only enable fair comparisons between methods but also provide a reliable foundation for developing robust models capable of handling real-world diversity, while unnecessary variability is reduced through harmonization. Therefore, addressing dataset heterogeneity and ensuring access to standardized, curated, and bias-aware resources is essential for advancing the clinical applicability of AI models.

Some studies have explored harmonization and pre-processing techniques for mammography images; however, these efforts have primarily focused on contrast enhancement methods aimed at improving the visibility of abnormalities or anatomical structures \cite{deng2016mammogram, perez2020deep, cao2021breast, perre2017influence, seoni2024all}. Notably, most of these techniques require parameter adjustments, which may vary across clinical environments and patient populations, thereby limiting their generalizability. Other critical technical aspects of harmonization are often overlooked or only vaguely addressed in existing models. \cite{kilintzis2024public} proposed a framework for harmonizing breast cancer datasets from five public repositories, but their work was primarily focused on Magnetic Resonance Imaging (MRI), leaving mammography-specific challenges insufficiently addressed. Recently, vision–language models have been proposed to leverage heterogeneous clinical data by pairing mammography images with radiology reports to improve breast cancer detection, even generating synthetic reports when textual data are unavailable from the provided abnormality annotations \cite{ghosh2024mammo}. However, our work takes a complementary direction by directly addressing the underlying issue of dataset heterogeneity and bias.

This paper provides a practical overview tailored specifically to mammography datasets. We begin by introducing key terminologies to establish a clear understanding of the datasets, their inherent abnormalities, and the nature of these variations. Building on this, we review available datasets, focusing on those that are publicly accessible or obtainable upon request, and analyze their characteristics in detail. A critical challenge facing current mammography AI models is their limited generalization to unseen datasets. We hypothesize this stems from quantifiable dataset biases. To address this, we present \textit{MammoClean}, a harmonization framework designed to mitigate these biases rather than merely serve as a pre-processing tool. Applied to three public datasets, the pipeline includes detailed technical considerations for robust dataset preparation and is released publicly to enhance community adoption and reproducibility. Finally, we conduct evaluations to identify and quantify bias sources that could undermine model performance across different datasets or in real-world clinical settings.

This paper may be particularly useful for researchers and developers in medical imaging, as well as for clinicians and dataset curators. For researchers and AI model developers, it offers perspectives on the intrinsic properties of mammography images and their associated abnormalities, which could support the design of models that are more robust to population-specific differences and less sensitive to case-selection biases. Clinicians might gain insights into how imaging variations and dataset composition can influence the diagnostic performance of AI models, even in ways that may not directly affect radiologists. Dataset curators and organizers may also find suggestions on what types of additional information could be considered in future datasets to enhance their research and clinical relevance. By attempting to bridge technical and clinical perspectives, this work aims to contribute to the ongoing efforts toward more reliable, fair, and clinically meaningful AI-driven solutions for breast cancer detection and diagnosis. Our contributions are summarized as follows:

\begin{itemize}
    \item We provide a comprehensive review of mammography imaging fundamentals. This includes breast anatomy, tissue texture characteristics, and detailed analysis of main findings and their radiological presentations. We follow this with an in-depth survey of publicly available mammography datasets. This survey reveals substantial heterogeneity in their characteristics, metadata structures, and annotation styles.
    \item We propose \textit{MammoClean}, a public pipeline that goes beyond traditional pre-processing for dataset harmonization. The pipeline offers three key capabilities. First, it provides extendibility through a modular, open-source architecture adaptable to new datasets. Second, it enables bias quantification that systematically identifies and measures dataset biases across clinical decisions, breast density distributions, and abnormality prevalence. Third, it ensures reproducibility through fully documented and standardized processing steps, enabling consistent dataset preparation across studies.
    \item We conduct a detailed evaluation of three harmonized public datasets, revealing key differences in clinical characteristics. It demonstrates how standardization facilitates fairer model comparisons and more robust development strategies for AI-driven mammographic image analysis. 
\end{itemize}

\section{Mammography Analysis}
In this section, we first describe breast anatomy and tissue composition as a key factor influencing both the visual appearance of mammography images and the associated risk and difficulty of abnormality detection. We then discuss the main categories of abnormalities typically observed in mammograms, followed by a brief overview of the clinical workflow.

\subsection{Breast Anatomy and Tissue Composition}

\begin{figure}
    \centering
    \includegraphics[width=0.95\linewidth]{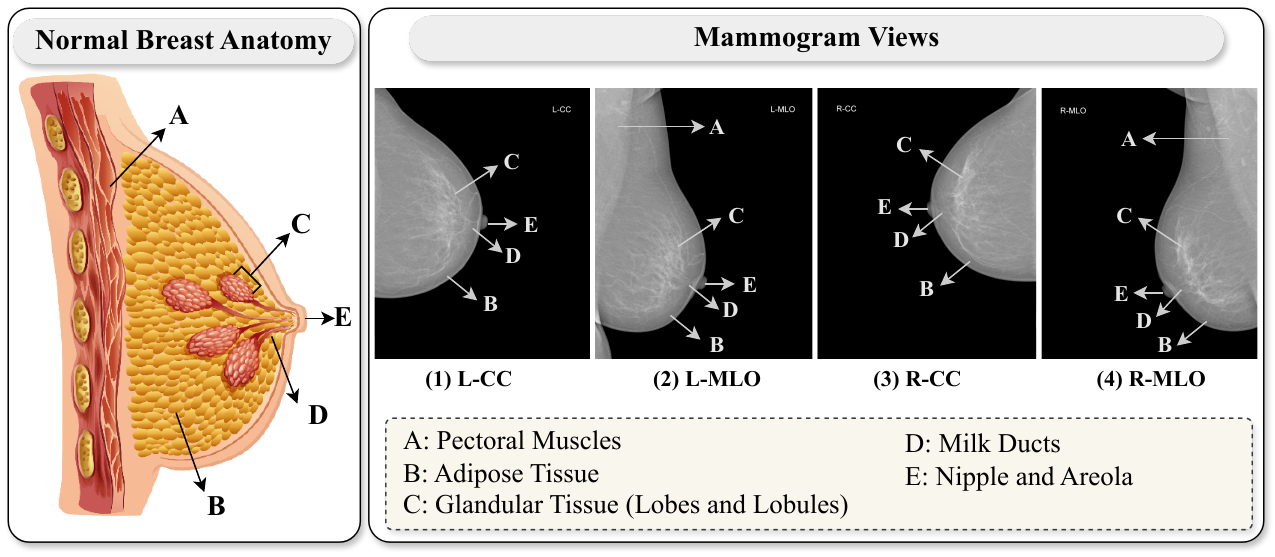}
    \caption{Illustration of breast anatomy and its appearance across different mammographic views, images are from \cite{nguyen2023vindr}.}
    \label{fig:breast_anatomy}
\end{figure}

The anatomy of the breast is a complex structure primarily composed of fibroglandular tissue (lobes and lobules), adipose tissue (fatty tissue), milk ducts, and blood and lymphatic vessels, all situated on the pectoral muscle. The fibroglandular tissue consists of several lobes, each containing numerous lobules that serve as the milk-producing units. Milk ducts are the channels that transport milk from the lobules to the nipple during lactation. Adipose tissue provides shape to the breast and fills the spaces around the lobes and ducts. Figure \ref{fig:breast_anatomy} illustrates the anatomy of the breast and its projections in different mammographic views.

The proportion of fibroglandular to adipose tissue varies among individuals, leading to different breast densities. Breast density plays a critical role in mammographic imaging, as it directly affects both the appearance of the image and the difficulty of detecting abnormalities, an aspect that must be efficiently learned by AI models. The Breast Imaging-Reporting and Data System (BI-RADS) provides a standardized four-category scale for classifying breast density based on the proportion of fibroglandular tissue \cite{spak2017bi}. Table \ref{tab:breast_density} outlines the details of these categories. See Figure \ref{fig:breast_density} for a visualization of the breast density categories and their visual differences as observed on mammography images. Importantly, both tumors and dense breast tissue exhibit similar radiographic intensities, making their differentiation particularly challenging for radiologists as well as for AI-based systems.

\begin{table}[ht]
\centering
\caption{ BI-RADS scaling for breast density, based on 5th edition \cite{spak2017bi}.}
\label{tab:breast_density}
\resizebox{0.5\textwidth}{!}{%
\begin{tabular}{|l|l|}
\hline
\textbf{BI-RADS } & \textbf{Type }  \\ \hline
A & Almost entirely fatty  \\ \hline
B & Scattered areas of fibroglandular density \\  \hline
C & Heterogeneously dense  \\ \hline
D & Extremely dense \\ \hline
\end{tabular} }
\end{table}

Breast density is influenced by several biological and demographic factors. One primary difference is observed between genders: men generally have denser breast tissue compared to women. However, this property is not static and changes dynamically over time. Breast density is inversely correlated with age; younger individuals typically have higher breast density, while older individuals exhibit lower density due to hormonal changes and the gradual replacement of fibroglandular tissue with adipose tissue \cite{checka2012relationship}. Another important factor affecting AI model performance across populations is ethnicity. Population-based studies have shown significant variation in breast density across ethnic groups; for instance, the prevalence of dense breast tissue is higher among Asian women compared to White women \cite{kerlikowske2023impact}.

\begin{figure}
    \centering
    \includegraphics[width=0.55\linewidth]{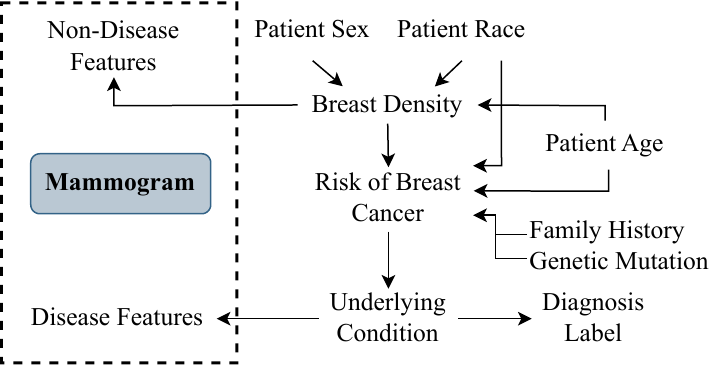}
    \caption{Causal relationships between various factors and their impact on mammography images for both disease-related and non-disease-related features. Patient gender, ethnicity, and age are common factors influencing breast density, which directly affects non-disease visual appearances in images and, by altering breast cancer risk and detection difficulty, also impacts disease-related features.}
    \label{fig:causal_relation}
\end{figure}

We extended the schematic proposed by \cite{jones2024causal} to illustrate the relationships among different influencing factors and their impact on both disease-related and non-disease-related imaging features based on age, gender, ethnicity, family history, and genetic mutation (see Fig. \ref{fig:causal_relation}). It is important to emphasize breast density is a key metadata component for AI model deployment. A robust dataset should avoid bias toward particular tissue types to ensure that developed models are both reliable and generalizable across diverse populations and imaging conditions.

\subsection{Common Mammographic Findings}

While interpreting the mammograms, existing visual cues assist radiologists for detecting potential abnormalities. These findings and their set of characteristics has an impact on the clinical decision making, such as its likelihood of malignancy or patient management approaches. Mammographic findings are generally classified into four main categories: masses, calcifications, architectural distortion, and asymmetry. Additional findings, such as skin thickening, skin retraction, nipple retraction, and other secondary signs, may also be observed. The primary categories are summarized as follows:

\textbf{Masses} are space-occupying lesions visibale in different mammogram projection. The analysis of a mass is based on three primary morphological features: shape, margin, and density \cite{baker1996breast, magny2023breast}. Benign masses, such as cysts or fibroadenomas, tend to be oval or round in shape with circumscribed (well-defined) margins and are often of low or equal density. On the other hand, malignant masses are frequently irregular in shape with spiculated, microlobulated, or indistinct margins. Comparing to the normal fibroglandur tissue, a higher density is also is high likely to be associated with malignancy. Detecting these patterns and characteristic for an AI model is a critical aspect that should not be over looked and the detection of these pattern in several subgropus must be examined to ensure the model's performance is not related to existing possible biases in datasets rather than actual underlying reasons. See Figure \ref{fig:mass_types} for a visualization of different mass types.

\textbf{Calcifications} are tiny calcified deposits within breast tissue that appear as bright dots in mammograms. They are characterized based on their morphology and distribution \cite{baker1996breast}. Typically benign calcifications include skin, vascular, and coarse calcifications, while suspicious ones are typically small ($<$0.5 mm) including fine linear and amorphous types. Regarding the distribution, clustered (grouped), linear, or segmental distributions are often suspicious. A robust AI-model must be able to localize this fine-grained patterns in high-resolution mammograoms which is a challenge due to computational costs and as typically most existing approaches resize the image to lower resolutions, these fine-details for small-size abnormalities may be lost during the under-sampling process. See Figure \ref{fig:calcification_types} for a visualization of different calcification types.

\textbf{Architectural Distortion} is defined as a disruption of the normal breast architecture without the presence of a discrete mass \cite{barazi2023mammography}. It may occur as a result of post-surgical or post-therapeutic changes; however, when no benign cause is identified, it is considered highly suspicious for malignancy and requires careful evaluation.

\textbf{Asymmetry} refers to an area of increased fibroglandular density that appears different when compared with the corresponding region in the contralateral breast. According to BI-RADS, there are four subtypes of asymmetry \cite{spak2017bi}. An asymmetry is a finding seen only in a single mammographic projection and is often difficult to characterize due to summation of normal tissue. A focal asymmetry is a localized area of density visible on at least two projections, but it lacks the borders and convex margins of a true mass. A global asymmetry involves a larger volume of tissue, typically occupying at least one quadrant of the breast, without the associated features of a suspicious lesion. A developing asymmetry is a new finding or one that has increased in size or conspicuity compared to prior examinations, and it is regarded as more clinically significant, often warranting additional diagnostic workup.

\subsection{Clinical Workflow}

In clinical practice, the mammographer specifies the purpose of the imaging study, whether it is performed for screening, diagnostic evaluation, or follow-up. The mammography report includes information on breast density as well as a detailed description of all observed findings with their associated characteristics. The final step involves the radiologist’s assessment, which is categorized using the BI-RADS scale, consisting of seven categories (see Table \ref{tab:birads}). Screening mammography results are typically limited to BI-RADS categories 0, 1, and 2, while the remaining categories are more relevant to diagnostic procedures. Following BI-RADS assessment, clinical management generally falls into one of four pathways: additional imaging, routine screening, short-term follow-up, or biopsy. For cases requiring biopsy, histopathological examination serves as the gold standard for determining malignancy and confirming the diagnosis of breast cancer.

\begin{table}[H]
\centering
\caption{BI-RADS scaling for mammography assessment and corresponding interpretations.}
\label{tab:birads}
\resizebox{\textwidth}{!}{%
\begin{tabular}{|l|l|}
\hline
\textbf{BI-RADS} & \textbf{Interpretation}  \\ \hline
0 & Incomplete assessment - Additional imaging needed and further evaluation required  \\ \hline
1 & Negative – No abnormalities detected \\  \hline
2 & Benign findings (e.g., cysts)  \\ \hline
3 & Probably benign (less than 2\% malignancy risk) and short-time follow-up is recommended \\ \hline
4 & Suspicious abnormality (2-95\% malignancy risk) and biopsy may be considered \\  \hline
5 & Highly suggestive of malignancy (more than 95\% risk) \\ \hline
6 & Known biopsy-proven malignancy that requires definitive treatment (surgery/chemotherapy) \\ \hline
\end{tabular} }
\end{table}

\section{Available Mammography Datasets}

Several datasets have been released for mammography image analysis, each containing specific types of information such as image annotations, breast density labels, or patient medical history. Some of these datasets are publicly available, while others are restricted. This section provides an overview of the major datasets and the information they include. A comparative summary of the available datasets is presented in Table \ref{tab:dataset_summary}. Studies that rely solely on private or proprietary datasets, which are not accessible to the research community, fall outside the scope of this work and have therefore been excluded.

\textbf{MIAS:} The Mammographic Image Analysis Society (MIAS) digital mammogram database consists of 322 8-bit digitized film-screen mammograms from 116 patients, all acquired in the MLO view \cite{suckling1994mammographic}. The images were collected from a single center in the United Kingdom and annotated by expert radiologists for breast density as well as for findings in the following categories: calcifications, well-defined/circumscribed masses, spiculated masses, other masses, architectural distortion, asymmetry, and normal cases. Each abnormality was approximately localized using a circular annotation, specified by the coordinates of the center and the corresponding radius (in pixels). In addition to lesion localization, severity and diagnostic assessments were provided, classifying cases into benign or malignant groups. Breast density was assigned to three categories, fatty, fatty-glandular and dense-glandular, and no patients from the extremely dense subgroup were included in this dataset.

\textbf{INBreast:} The INBreast dataset \cite{moreira2012inbreast} was collected in Portugal between 2008 and 2010 and consists of 14-bit FFDM images with resolutions ranging from 3328 $\times$ 4084 and 2560 $\times$ 3328 pixels. The dataset contains 115 cases originating from screening, diagnostic, and follow-up studies. For 90 cases, images of both breasts were provided, including CC and MLO views (a total of four images per case). For the remaining 25 cases, only two views of a single breast were available. In addition, follow-up examinations were included for 8 cases. All abnormalities were annotated by a specialist and verified by a second expert. Annotations were provided as detailed contours delineating calcifications, masses, asymmetries, distortions, and the pectoral muscle (only for MLO view). Additional metadata included patient age, medical reports, breast density assessments, and BI-RADS categories. For 56 cases with BI-RADS scores greater than 2, biopsy results were available: 11 were confirmed as benign and the remainder malignant. The remaining cases were considered benign without biopsy confirmation. However, biopsy results are not provided in the available version at the time. However, it should be noted that the biopsy results and the patient age are not included in the currently available public version of the dataset.

\textbf{DDSM and CBIS-DDSM:} The Digital Database for Screening Mammography (DDSM) \cite{heath1998current, heat2000digital} consists of 2,620 screen-film mammography studies, each containing four images, for a total of 10,480 images. The dataset includes normal, benign, and malignant cases, with associated metadata such as patient age, breast density, and BI-RADS assessment. Lesion annotations were provided as approximate regions of interest (ROIs), indicating the general location of abnormalities.

\begin{figure}
    \centering
    \includegraphics[width=0.95\linewidth]{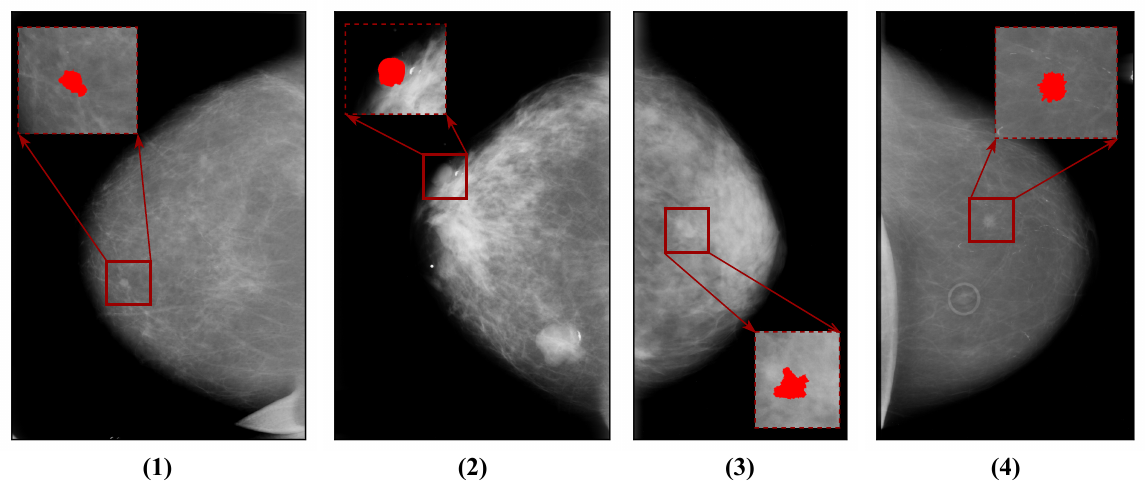}
    \caption{Different mass types, images are from \cite{song2009breast}: (1) round with circumscribed margins (BI-RADS 2, benign), (2) oval with circumscribed margins (BI-RADS 3, benign), (3) oval with ill-defined margins (BI-RADS 4, malignant), and (4) irregular shape with spiculated margins (BI-RADS 5, malignant).}
    \label{fig:mass_types}
\end{figure}

Due to limitations of the original dataset, including the relatively imprecise lesion annotations \cite{song2009breast} and the outdated SFM format, the Curated Breast Imaging Subset of DDSM (CBIS-DDSM) \cite{lee2017curated} was later developed. In this curated dataset, a subset of DDSM cases was re-annotated by specialized radiologists to improve annotation accuracy and to review questionable cases for lesion visibility. CBIS-DDSM includes 753 cases with calcifications and 891 cases with masses, with some overlap where both findings are present in the same case. The dataset is partitioned into training and test sets across four categories based on abnormality type (mass or calcification). Despite this data arrangement, asymmetry and distortion abnormalities were also provided in the mass group. Additional metadata provided in CBIS-DDSM includes the number of abnormalities per image, mass shape, mass margin, calcification type, calcification distribution, BI-RADS assessment, pathology results (benign, malignant, and benign without callback), and subtlety ratings for abnormality visibility. However, patient age information is not included in this subset. See Figure \ref{fig:mass_types} for some sample images from this dataset.

\textbf{CSAW-CC:} The Cohort of Screen-Aged Women Case-Control (CSAW-CC) \cite{dembrower2020multi} is a population-based study of women aged 40–74 years in Sweden. FFDM data were collected from three breast centers between 2008 and 2015, resulting in a dataset of 499,807 women with a total of 1,182,733 images. The complete dataset is not publicly available, but a restricted subset can be accessed, which includes 8,723 women, of whom 873 were diagnosed with breast cancer during the observation period while the remaining served as controls. Most participants underwent more than one examination, and no examinations performed after diagnosis were included. For cancer-diagnosed cases, all prior examinations were assigned the same label, with the time between screening and diagnosis reported in three categories: less than 60 days from screening to diagnosis corresponding to screen-detected cases, 60 to 729 days corresponding to interval cancers, and more than 730 days corresponding to prior studies. 

Other available information in the dataset includes cancer laterality for patients diagnosed with breast cancer, as well as cancer type, which is categorized into three groups: in situ, invasive tumors smaller than 15 mm, and invasive tumors larger than 15 mm. Patient age is provided in two categories, 40–55 years and older than 55 years, and lymph node status is reported as a binary variable indicating the presence or absence of metastasis. An important feature of this dataset is the documentation of radiologists’ decisions, including whether a subject was considered healthy, required further discussion, or was recalled, along with the assigned recall label. Additional information is also available regarding the reason for recall, distinguishing between cases based on radiological findings in the images and those prompted by clinical symptoms.

\textbf{CMMD and TOMPEI-CMMD:} The Cancer Mammography Database (CMMD) \cite{cui2021chinese} is a large-scale dataset consisting of FFDM images from 1,775 patients collected from 2012 to 2016. While some patients have all four standard views, others include only two views from a single breast. The available metadata includes biopsy results with cancer subtype information, the type of abnormality (mass, calcification, or both), and patient age. However, the dataset does not provide annotations regarding the location of abnormalities or their morphological characteristics.

\begin{figure}
    \centering
    \includegraphics[width=0.95\linewidth]{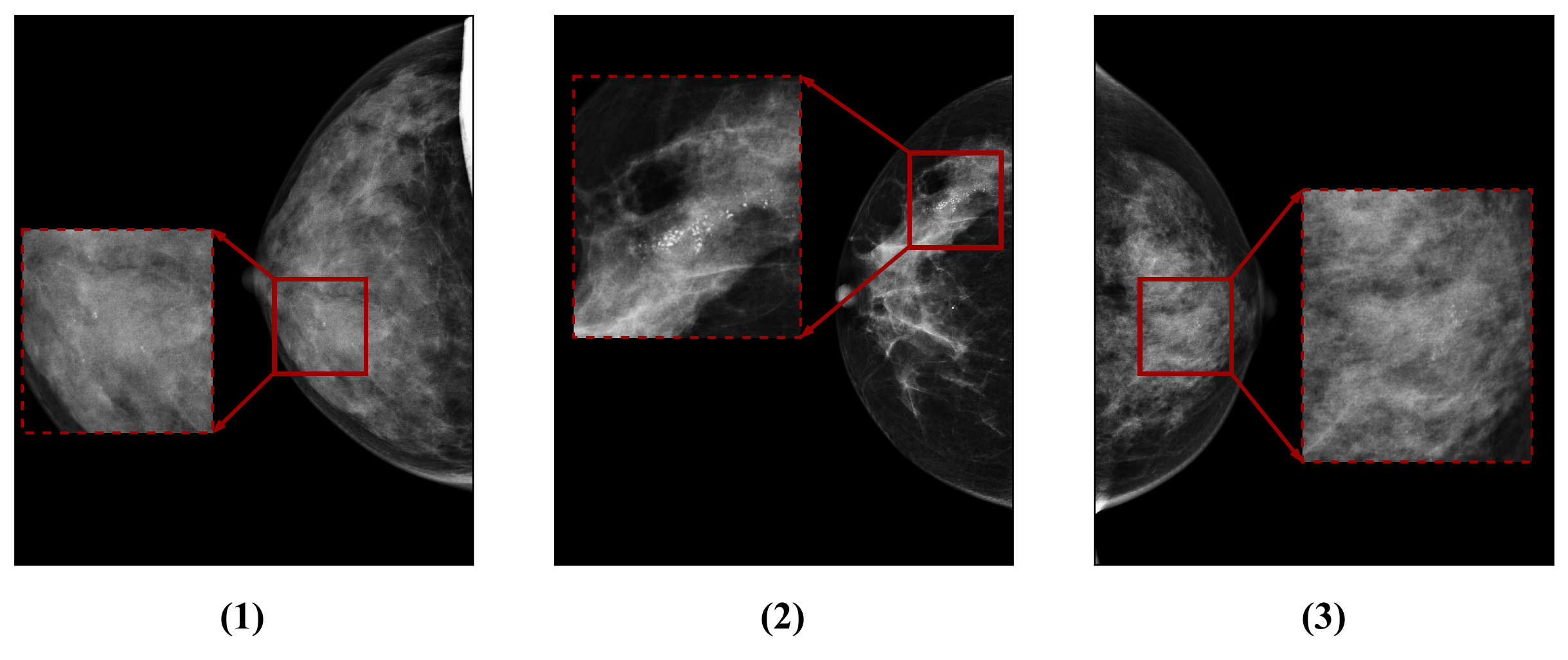}
    \caption{Different calcification types, images are from \cite{cui2021chinese, kashiwada2025tompei}: (1) small round calcifications with scattered distribution throughout the whole breast (BI-RADS 2, benign), (2) pleomorphic calcifications with segmental distribution in the medial breast (BI-RADS 5, malignant), and (3) amorphous, indistinct calcifications with grouped distribution in the medial breast (BI-RADS 3, malignant).}
    \label{fig:calcification_types}
\end{figure}

The updated version, TOMPEI-CMMD \cite{kashiwada2025tompei}, addresses several limitations of the original dataset and introduces additional annotations. Labeling errors involving laterality (right/left breast) and view type (MLO/CC) were corrected to improve dataset accuracy and some cases with non-visible lesions were excluded. For cases with masses or calcifications, both lesion characteristics and locations were annotated, along with the number of abnormalities present. Furthermore, additional abnormality types such as asymmetry and architectural distortion were included. Lesion locations were annotated in a descriptive, breast-based manner, categorized as lower region, medial side, middle to lateral, subareolar region, upper region, upper-medial region, and entire region. BI-RADS assessments and breast density information were also added. See Figure \ref{fig:calcification_types} for some sample images from this dataset.

\textbf{KAU-BCMD:} The King Abdulaziz University Breast Cancer Mammogram Dataset (KAU-BCMD) \cite{alsolami2021king} was collected between 2019 and 2020 and includes 1,416 cases, each annotated with BI-RADS scores that were reviewed by three radiologists. For 205 of these cases, corresponding ultrasound (US) images are also available, similarly annotated with BI-RADS assessments. While BI-RADS category 2 represents the largest group in the metadata, the currently available public version of the dataset does not include mammograms from BI-RADS 2 cases or the ultrasound images.

\textbf{RSNA:} This dataset was collected from sites in the United States and Australia and was released as part of a Kaggle competition \cite{carr2022rsna}. It is divided into training and test subsets, with the test set containing limited metadata. The dataset is based on screening FFDM images, with most cases including all four standard views, although some have missing views. Biopsy-verified labels for cancer presence are provided, and in cases of confirmed cancer, information regarding invasive versus non-invasive subtype is also included. Since the dataset originates from screening examinations, BI-RADS scores are limited to categories 0, 1, and 2, with higher categories not represented.

\begin{figure}
    \centering
    \includegraphics[width=1\linewidth]{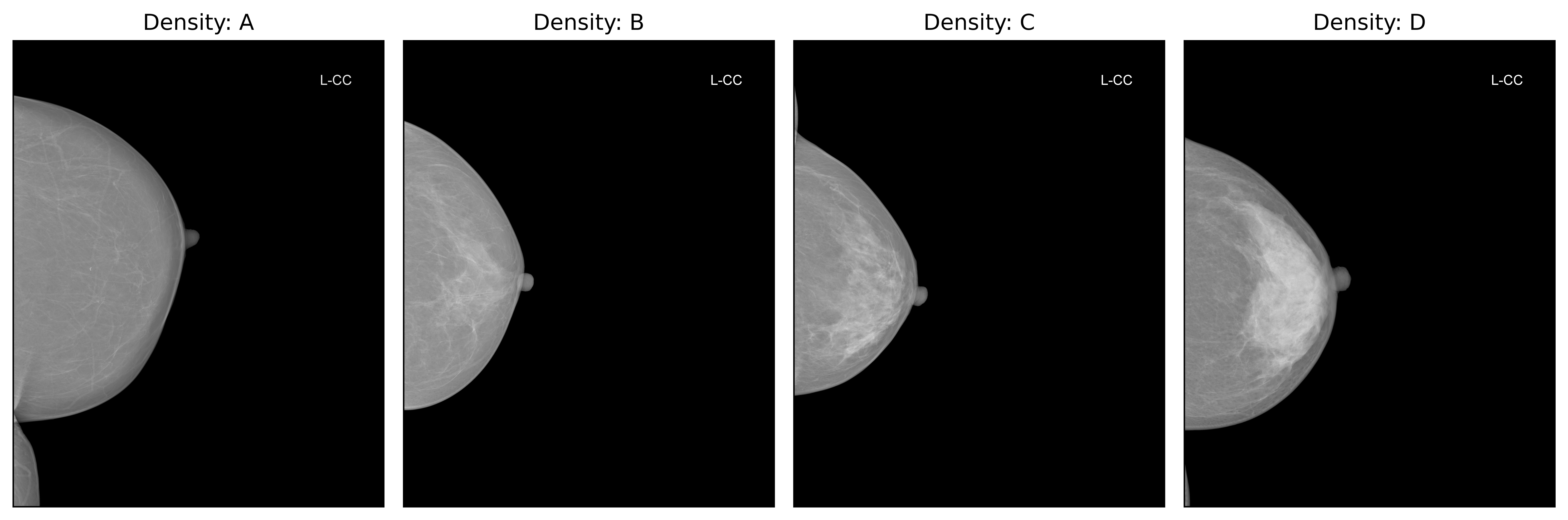}
    \caption{Different breast density categories on left CC mammography images from normal cases, images are from \cite{nguyen2023vindr} (A: almost entirely fatty, B: scattered areas, C: heterogeneously dense, and D: extremely dense).}
    \label{fig:breast_density}
\end{figure}

\textbf{VinDr-Mammo:} This dataset was collected from two primary hospitals in Hanoi, Vietnam \cite{nguyen2023vindr} and contains four-view FFDM images from 5,000 patients. See Figure \ref{fig:breast_density} for some sample images from this dataset. All images were independently reviewed by two radiologists, and in cases of disagreement, a third radiologist provided the final decision. The annotated findings include masses, suspicious calcifications, asymmetry, focal asymmetry, global asymmetry, architectural distortion, skin thickening, skin retraction, nipple retraction, and suspicious lymph nodes. Each breast image was assigned a BI-RADS score and a breast density category. For non-benign findings (BI-RADS score $>$ 2), bounding box coordinates were provided to localize the abnormality. It is important to note that biopsy-confirmed pathology results and detailed characteristic of findings such as mass and calcification are not available in this dataset.

\textbf{MMD:} The Mammogram Mastery Dataset (MMD) \cite{aqdar2024mammogram} was collected from four clinical units in Iraq and contains mammography images from 745 patients. For each case, either a CC or MLO view was provided, and the dataset was divided into two groups: cancer and non-cancer. In addition to the original collection, an augmented version of the dataset was released, generated using multiple image transformations such as rotation, flipping, affine transformations, noise addition, and elastic deformation. No additional clinical information or annotations were included with the dataset. It is important to note that while augmentation can improve data diversity for training purposes, the use of elastic transformation may alter key visual features that serve as indicators of malignancy, and therefore, the augmented version should be employed with caution.

\textbf{EMBED:} The Emory Breast Imaging Dataset (EMBED) \cite{jeong2023emory} was collected between 2013 and 2020 from four institutions in the United States. It includes all women aged over 18 years who had at least one mammogram archived during this period, resulting in a total of 115,910 patients and 3,383,659 images. In addition, approximately 40,000 linked ROI annotations for lesions are provided. One of the key aims of developing this dataset was to address the limitations of existing resources regarding race and ethnicity imbalance, and it therefore contains a more diverse population distribution, representative of the patient demographics across the contributing institutions.

The dataset comprises both screening and diagnostic studies, though the majority of the ROI annotations are associated with screening examinations. Images include FFDM, DBT, and synthesized 2D mammography, with about 42\% of the examinations containing both FFDM and DBT. A comprehensive set of imaging descriptors has been applied to characterize findings, including mass (shape, margin, density), calcifications (morphology and distribution), and other abnormalities, along with detailed size and positional information. Pathology reports were also used to categorize abnormalities into seven groups reflecting severity: invasive cancer, in-situ (non-invasive) cancer, high-risk lesions, borderline lesions, non-breast cancers, and normal findings. 

\textbf{OMI-DB:} The Optimam Mammography Image Database (OMI-DB) is a large-scale database that was collected from over 465,000 women across multiple sites in the United Kingdom and includes 1,311,413 studies with a total of 6,998,298 processed and unprocessed mammography images, spanning more than 10 years beginning in 2008 \cite{halling2020optimam}. Each case contains FFDM, DBT, or both, and for a subset of high-risk cases, MRI is also available. Images are labeled as malignant, benign, interval, or normal. In addition to clinical, surgical, and pathological data available for all cases, a subset of images includes expert-provided ROI annotations. Longitudinal data and temporal linkage are also available for patients who underwent repeated studies during the collection period. The dataset includes images acquired from a range of imaging devices, including those from Hologic, Siemens, GE Healthcare, and Philips. Additional metadata includes cancer type (invasive or in-situ) and patient age. Although the dataset is accessible under restricted conditions, certain information, such as breast density, requires additional approval.

\textbf{CDD-CESM:} This dataset was collected in Egypt and consists of contrast-enhanced spectral mammography (CESM) images from 326 patients aged 18 to 90 years \cite{khaled2021categorized}. CESM was performed using standard mammography equipment following the injection of a contrast agent, with images acquired at two different energy levels. The low-energy images closely resemble conventional digital mammography, whereas the high-energy exposures capture functional activity based on contrast enhancement. For each patient, both the low-energy images and the subtracted images (where abnormalities are more clearly highlighted) are provided. Most patients have eight images available; however, for some patients, only four images from a single breast are included. In certain cases, views are missing due to low image quality or unavailability. In addition to imaging data, the dataset provides clinical information such as patient age, breast density, biopsy-verified results, BI-RADS assessments, abnormality characteristics, and radiology reports.

\textbf{BCS-DBT:} This dataset comprises 22,032 DBT images from 5,060 patients, collected at the Duke Health System between 2014 and 2018. It includes a total of 5,610 studies, with some patients contributing more than one study \cite{buda2021data}. Each study contains at least one view, though the number of views per study varies. The dataset was organized into four groups based on BI-RADS assessments and pathology outcomes: (1) normal studies, defined as those with BI-RADS 1 assessments in radiology reports; (2) actionable studies, in which radiologists recommended additional imaging; (3) benign studies, selected from BI-RADS 4 assessments that underwent biopsy and were confirmed benign; and (4) cancer studies, selected from BI-RADS 4 and 5 assessments that underwent biopsy and were confirmed malignant. For the actionable, benign, and cancer groups, the initial case selection was limited to those with reported mass or architectural distortion abnormalities, while calcification cases were excluded due to their distinct visual characteristics.

\begin{sidewaystable}[ph!]
\centering
\renewcommand{\arraystretch}{2}
\resizebox{\textheight}{!}{%
\begin{tabular}{l l l l l l l p{6cm} p{4cm} l l l}
\toprule
Dataset & Year & Country & \# Studies (Longitudinal) & \# Images & Image Acq. Mode & Diagnosis & Finding Types & Annotations & BI-RADS Score & Breast Density & Age \\
\midrule

MIAS \cite{suckling1994mammographic} & 1994 & United Kingdom & 161 ($\times$) & 332 & SFM & $\checkmark$ & Normal, Mass, Calcification, Distortion, Asymmetry & Circle around findings with coordinates of center and approximate radius & $\times$ & $\checkmark$ & $\times$ \\

CBIS-DDSM \cite{lee2017curated} & 2017 (1999$^{+}$) & United States & 1,391 ($\times$) & 2,844 & SFM & $\checkmark$ & Mass, Calcification, Distortion, Asymmetry with characteristics & Contours enclosing the findings & $\checkmark$ & $\checkmark$ & $\times$ \\

InBreast \cite{moreira2012inbreast} & 2012 & Portugal & 115 (Only 8) & 410 & FFDM & $\times$ & Normal, Mass, Calcification, Distortion, Asymmetry & Contours enclosing findings  & $\checkmark$ & $\checkmark$ & $\times$ \\

TOMPEI-CMMD \cite{kashiwada2025tompei} & 2025 (2021$^{+}$) & China & 1,775  ($\times$) & 3,728  & FFDM & $\checkmark$ & Normal, Mass, Calcification, Distortion, Asymmetry, and other features with characteristics & Descriptive anatomical region annotation & $\checkmark$ & $\checkmark$ & $\checkmark$ \\

CSAW-CC \cite{dembrower2020multi}$^{*}$ & 2020 & Sweden & 8,723 ($\checkmark$) & 98,788  & FFDM & $\checkmark$ & Normal, Invasive/Non-Invasive Cancer & Segmentation mask for some of the screen detected cases & $\times$ & $\times$ & Categorical \\

KAU-BCMD \cite{alsolami2021king} & 2021 & Saudi Arabia & 1,416 ($\times$) & 5,687 & FFDM, US & $\times$ & N/A & N/A & $\checkmark$  & $\times$  & $\checkmark$  \\

RSNA \cite{carr2022rsna} & 2022 & United States, Australia & 1,970 ($\times$) & 9,594 & FFDM & $\checkmark$ & Normal, Invasive Cancer, Non-Invasive Cancer & N/A & $\checkmark$ & $\checkmark$ & $\checkmark$ \\

VinDr-Mammo \cite{nguyen2023vindr} & 2022 & Vietnam & 5,000 ($\times$) & 20,000 & FFDM & $\times$ & Normal, Mass, Calcification, Distortion, Asymmetry, and other features & Bounding box around findings for BI-RADS higher than 2 & $\checkmark$ & $\checkmark$ & $\checkmark$ \\

MMD \cite{aqdar2024mammogram} & 2024 & Iraq & 745 ($\times$) & 745 & FFDM & $\checkmark$  & N/A  & N/A & $\times$ & $\times$ & $\times$ \\

EMBED \cite{jeong2023emory}$^{*}$ & 2023 & United States & 115,910 ($\checkmark$) & 3,383,659 & FFDM, DBT & $\checkmark$ & Normal, Mass, Calcification, Distortion, Asymmetry, Invasive Cancer, Non-Invasive Cancer, Non-Breast Cancer, High-Risk Lesion, Borderline Lesion & Region of interest & $\checkmark$ & $\times$ & $\checkmark$ \\

OMI-DB \cite{halling2020optimam}$^{*}$ & 2020 & United Kingdom & 1,311,413 ($\checkmark$) & 6,998,298 & FFDM, DBT, MRI & $\checkmark$ & Normal, Invasive Cancer, Non-invasive Cancer & Region of interest & N/A & $\checkmark$ & $\checkmark$ \\

CDD-CESM \cite{khaled2021categorized} & 2021 & Egypt & 326 ($\times$) & 2,006 & CESM  & $\checkmark$ & Normal, Mass, Calcification, Distortion, Asymmetry & Segmentation masks & $\checkmark$ & $\checkmark$ & $\checkmark$ \\

BCS-DBT \cite{buda2021data} & 2021 & United States & 5,610 ($\checkmark$ ) & 22,032 & DBT & $\checkmark$ & N/A & Bounding box enclosing findings & $\times$ & $\times$ & $\times$ \\
\bottomrule
\end{tabular}}
\caption{Summary of publicly available mammography datasets. $^{*}$ indicates datasets with restricted access, and $^{+}$ denotes the year datasets initially published with a later curated version providing more precise and detailed information. }
\label{tab:dataset_summary}
\end{sidewaystable}

\section{Harmonizing Public Datasets}

We developed a public-access pipeline, \textbf{MammoClean}, designed for processing and harmonizing mammography datasets. To evaluate its performance and investigate potential data quality issues and sources of heterogeneity, we applied the pipeline to three publicly available datasets: CBIS-DDSM, TOMPEI-CMMD, and VinDr-Mammo. The selection of these datasets was based on three criteria: open accessibility, sufficient number of studies and images, and the availability of annotations or characterization of findings that enable further evaluation of models trained on them. Although the current implementation focuses on these three datasets, \textit{MammoClean} can be extended and adapted for use with other mammography datasets.

Notably, \textit{MammoClean} was developed for multi-view image analysis, which has gained increasing attention in recent years as it provides a more comprehensive approach compared to single-view analysis \cite{zafari2025multi}. This design choice influences case selection and data management, which may differ from pipelines developed for single-view settings. The output of \textit{MammoClean} was defined to include breast density classification, BI-RADS assessment, and diagnostic labels, together with relevant clinical data such as patient age, where available. Additional information, including annotations of findings and their characteristics, can be easily retrieved from the metadata if needed.

\subsection{Common Issues and Sources of Heterogeneity}

Several factors contribute to heterogeneity in mammography datasets, which can significantly affect both the development and performance of deep learning models. Some of these sources of variation need to be mitigated through pre-processing or harmonization, while others should be explicitly preserved to ensure that models achieve robust performance across different clinical settings.

\textbf{Annotation inconsistency:} The annotations and descriptive information provided across different datasets vary considerably. While some can be converted into a common format, others are not directly interoperable. For example, the VinDr-Mammo dataset provides rectangular bounding-box annotations, whereas CBIS-DDSM includes detailed enclosing contours. Although bounding boxes can be generated from contours, the reverse is not possible, as detailed contours cannot be reconstructed from rectangular boxes. In contrast, TOMPEI-CMMD offers breast-level descriptors for abnormality locations, which cannot be converted to either bounding boxes or contours. This inconsistency poses a significant limitation for the development and evaluation of advanced models, particularly those designed for localization or segmentation tasks across multiple datasets.

Another example of inconsistency is the representation of breast density. VinDr-Mammo follows the BI-RADS standard, TOMPEI-CMMD records textual descriptions of density categories, while CBIS-DDSM stores them as numerical codes. Such discrepancies can be mitigated by selecting a standard framework and mapping all dataset-specific information accordingly, thereby improving harmonization.

\textbf{Bit Depth and dynamic range differences:} Mammography images are often stored in high bit depth to preserve subtle contrast variations that are critical for detecting small abnormalities. However, differences in acquisition protocols and storage formats result in inconsistencies in bit depth and dynamic range across datasets, which can complicate model training and evaluation. For example, the VinDr-Mammo dataset provides images with 16-bit depth, whereas the TOMPEI-CMMD dataset stores images in 8-bit depth. Such variability can lead to loss of diagnostic information and introduce dataset-specific biases. Therefore, addressing these differences through pre-processing techniques such as normalization is an essential step in dataset preparation for AI model development.

\textbf{Resolution variability:} Mammography images are inherently high-resolution, but their resolution can vary depending on the imaging device, acquisition view, and breast size. Most deep learning models require a fixed input resolution, which is typically achieved by resampling the images. However, naive resampling without appropriate consideration may distort the appearance of abnormalities, thereby affecting model performance. For instance, since different mammographic views can naturally have different resolutions, resampling them all to the same resolution without zero-padding can disproportionately distort one view compared to another, leading to incorrect relative sizes of abnormalities across views. Importantly, resolution variability is not limited to differences between datasets but can also occur within a single dataset. For example, the VinDr-Mammo dataset contains images with 58 different resolutions, ranging from $2812\times2012$ to $3580\times2812$. Noteworthy that resampling the images to lower resolution can lead to missing information of subtle abnormalities. 

\textbf{Laterality flipping:} Some datasets suffer from laterality inconsistencies, where the laterality information in the image headers does not match the actual side of the breast \cite{buda2021data}, or the laterality is correct but the image itself has been horizontally flipped. Ensuring consistent and accurate laterality annotation is a critical step in dataset preparation, particularly for four-view image analysis. A common pre-processing approach involves horizontally flipping one breast side to achieve consistency and facilitate ipsi-lateral comparisons, thereby improving feature extraction \cite{isosalo2023independent, manigrasso2025mammography, zafari2025hybrid}. In certain datasets these issues have been documented, whereas in others they require careful verification. See Figure \ref{fig:mass_types}, where all images are from the R-CC view but exhibit different laterality, illustrating this issue. 

\begin{figure}
    \centering
    \includegraphics[width=1\linewidth]{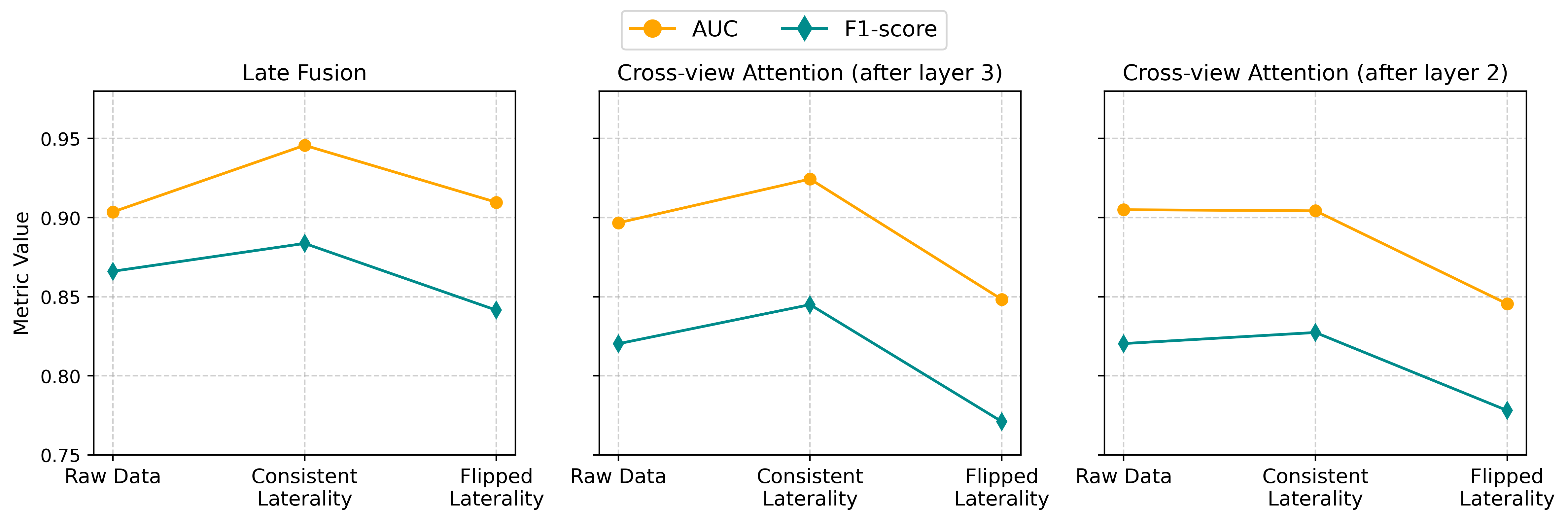}
    \caption{ Evaluation of ResNet18-based models with different fusion strategies using raw data, data with uniform laterality, and data affected by the flipped laterality issue.  }
    \label{fig:laterality_impact}
\end{figure}

To evaluate the impact of flipped laterality on deep learning models, we conducted a study comparing different multi-view fusion strategies using the ResNet18 \cite{he2016deep} architecture, one of the most widely adopted backbones for feature extraction in mammography analysis \cite{zafari2025multi, zafari2025hybrid}. We used a subset of the TOMPEI-CMMD dataset that included all four standard mammographic views and performed nine experiments based on varying fusion strategies and laterality conditions for binary malignancy classification. The dataset was organized under three laterality conditions: (1) \textit{Raw Data}, where left and right breast views retained their original orientations but shared consistent positioning within each laterality; (2) \textit{Consistent Laterality}, where right breast images were horizontally flipped to achieve uniform orientation across all views; and (3) \textit{Flipped Laterality}, where a random subset of images was horizontally flipped to simulate laterality inconsistency. The fusion strategies were defined as follows: (1) Late Fusion, where feature maps extracted by ResNet18 were combined at a later stage, following the approach of \cite{zafari2025hybrid}; (2) Cross-View Attention (Intermediate Fusion), where cross-view attention modules were inserted after layer 3 of ResNet18 to integrate information across both ipsi-lateral and contra-lateral views, inspired by \cite{van2021multi}; and (3) Cross-View Attention (Early-Level), where the attention module was applied after layer 2 to capture more subtle inter-view relationships, restricted to contra-lateral analysis due to computational constraints. The results of these experiments are presented in Figure~\ref{fig:laterality_impact}.

In nearly all evaluations, the presence of flipped laterality led to a decline in model performance. Comparing the results obtained from raw data with those from consistently oriented data revealed that maintaining uniform laterality improved model performance. These findings emphasize the importance of preserving a structure for image orientation, as inconsistencies, whether originating from dataset construction or data augmentation techniques, can adversely affect the learning process and overall performance of deep models. Automated detection of laterality inconsistencies can be facilitated by selecting subsets of the image from both sides and analyzing their intensity variance to identify discrepancies. Let be a mammography image of size $H \times W$, and let $n$ denote the window width used to extract a subset of the image from each lateral side. The laterality can then be determined by comparing the intensity variance (or standard deviation) of the left and right windows as follows:

\begin{equation}
\text{laterality} = 
\begin{cases}
\text{``left''} & \text{if } \sigma({I}[:, 0:n]) > \sigma({I}[:, W-n:W]) \\
\text{``right''} & \text{otherwise}
\end{cases}
\end{equation}

\textbf{Flipped intensity and background corruption:} Standard mammography images typically have a zero-valued background, with breast structures appearing brighter relative to it. However, in some datasets this convention is not preserved, and the background may instead have the highest intensity while the breast tissue appears darker, a problem known as flipped intensity. In certain cases, flipped intensity occurs together with corrupted background sampling, further complicating image interpretation for deep models. This issue is particularly common in the VinDr-Mammo dataset and detecting and correcting such intensity abnormalities is an important step in dataset pre-processing. 

\begin{figure}
    \centering
    \includegraphics[width=0.8\linewidth]{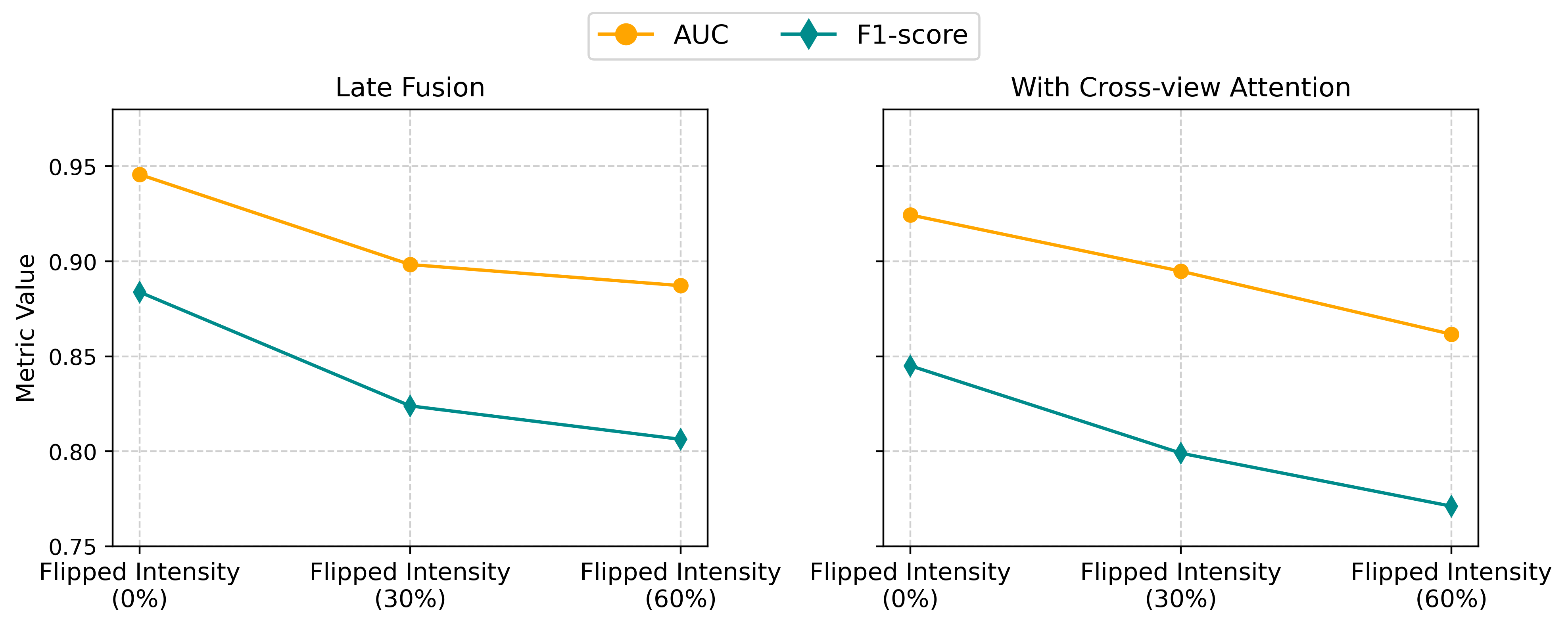}
    \caption{ Evaluation of ResNet18-based models' performance under varying percentages of data with flipped intensity.  }
    \label{fig:intensity_impact}
\end{figure}

To assess the effect of intensity-flipping artifacts on deep model performance, we employed the same dataset and architectures, randomly inverting image intensities for 30\% and 60\% of patients. The results (see Figure~\ref{fig:intensity_impact}) demonstrate that increasing the proportion of intensity-flipped images leads to a noticeable degradation in model performance, highlighting the sensitivity of deep networks to intensity inconsistencies in mammography data. 

This issue can be detected by comparing intensity distributions in two subsets of the image, similar to the approach used for laterality detection, where the statistical features of intensity values of predefined windows are compared. However, this method requires accurate knowledge of breast laterality to avoid misclassification. Alternative strategies, such as analyzing the image histogram, can also be applied, but these may be less robust due to variations in breast size or inter-dataset contrast differences. 

Several additional factors also contribute to dataset heterogeneity, including inconsistencies in file naming and organization, missing data or metadata, differences in acquisition protocols, variability across vendors and imaging devices, and population diversity. Data curation and management practices further amplify these challenges, as some datasets are organized at the patient level while others are structured at the image level, making harmonization and unified management more complex. Addressing these issues requires dataset-specific evaluations to ensure consistency before developing a standardized framework for deep model training and validation. Missing information, such as absent views or incomplete metadata, must also be carefully managed, either by excluding incomplete studies or by adopting relevant strategies that enable model development despite missing data \cite{zafari2025hybrid, park2025multi, germani2025bias}.

Image acquisition protocols, as well as variability across vendors and imaging devices, are additional factors that noticeably influence the visual appearance of mammography images, particularly in terms of contrast. Such shifts in contrast distribution are not only observed across datasets but also expected in real-world applications, where models must learn clinically relevant features rather than relying on dataset-specific visual patterns. Image enhancement techniques can help harmonize images from different sources by aligning their contrast distributions; however, these methods must be evaluated at large scale and remain flexible enough to adapt to diverse imaging conditions. Beyond technical variability, population diversity also plays a role in imaging characteristics; for example, breast density distributions may differ across populations, influencing the prevalence of certain conditions. Therefore, AI models should be designed and validated to perform robustly across both clinical and non-clinical sources of variability to ensure their adaptability and reliability in real-world workflows.

\subsection{Data Standardization and Preparation}
\begin{figure}
    \centering
    \includegraphics[width=1\linewidth]{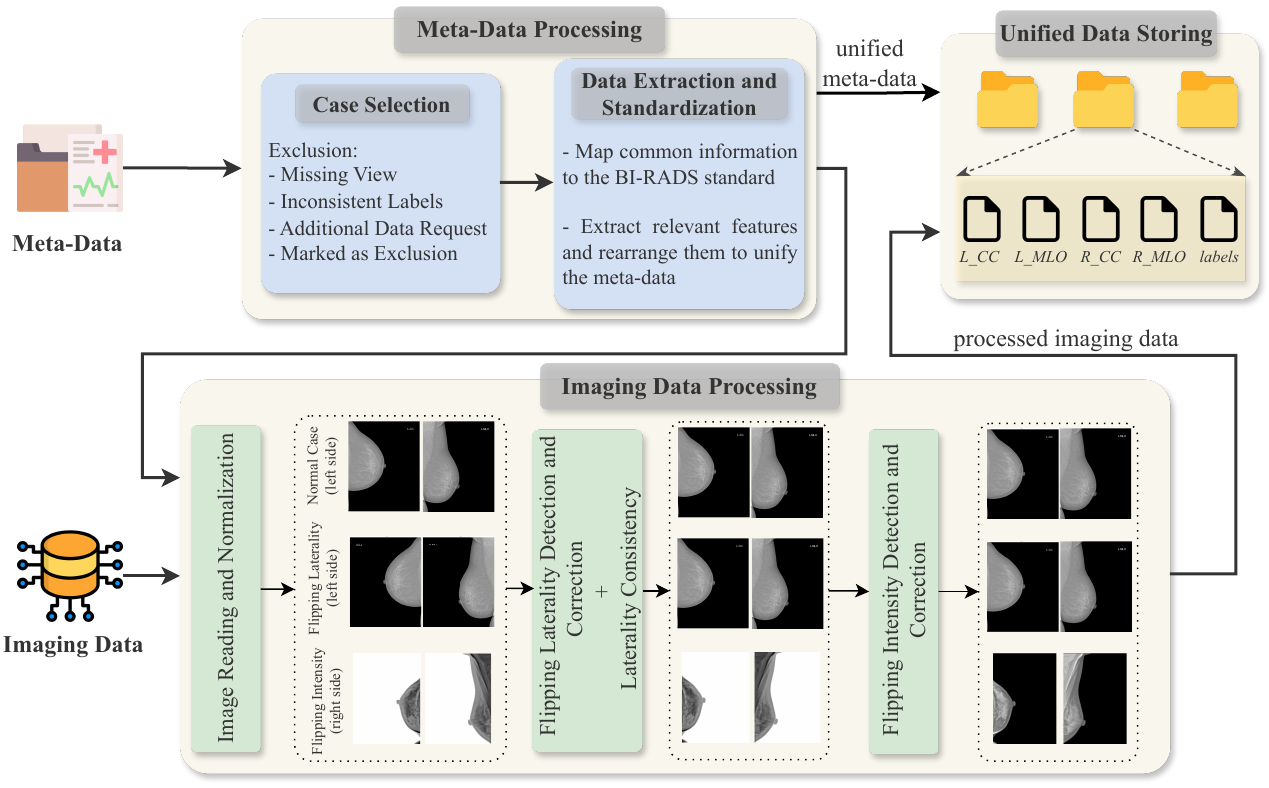}
    \caption{Illustration of the MammoClean process for pre-processing and storing mammography images and corresponding meta-data files.   }
    \label{fig:MammoClean}
\end{figure}

The process of data standardization and harmonization for deep learning models can be structured into three main stages: (1) case-selection and standardizing metadata, (2) pre-processing imaging data, and (3) storing data in a unified format. Figure \ref{fig:MammoClean} illustrates the meta-data and imaging data process and storing in \textit{MammoClean}. 

\subsubsection{Initial Case Selection}
\begin{figure}
    \centering
    \includegraphics[width=0.95\linewidth]{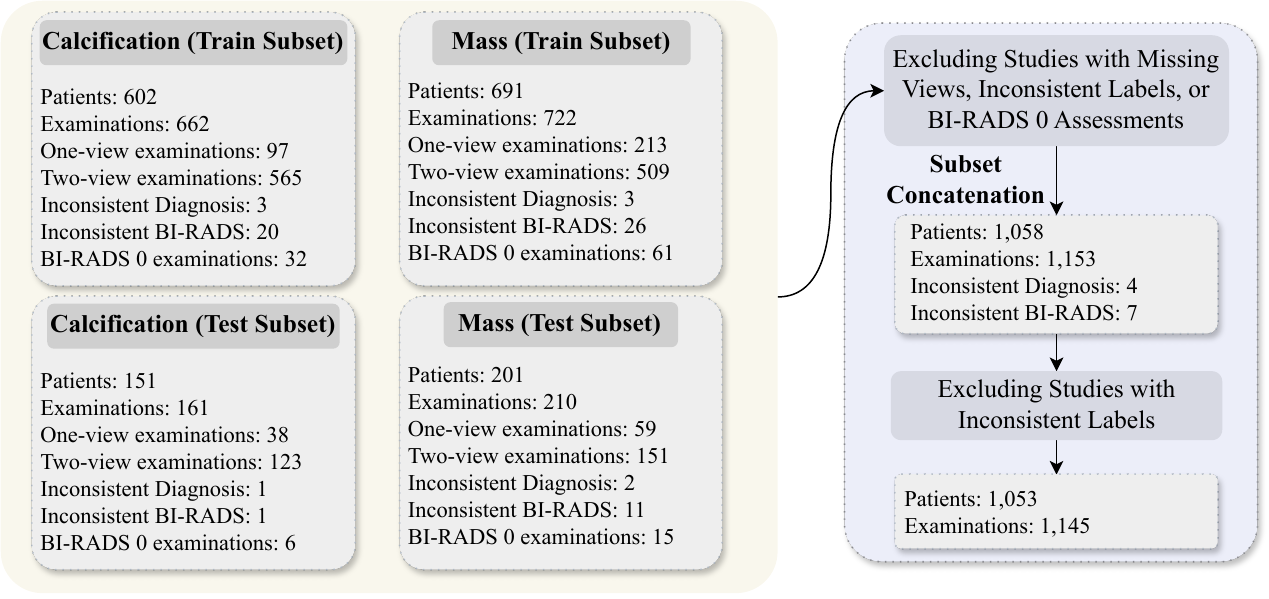}
    \caption{The process of case selection from the CBIS-DDSM dataset. Different subsets of images provided by the dataset were evaluated separately and then concatenated to assess possible inconsistencies. Examinations were defined on a breast basis, and those with missing views, inconsistent diagnosis labels or BI-RADS assessments, or BI-RADS 0 (requiring further imaging) were excluded. After concatenating data from different subsets, labels were double-checked to remove inconsistencies across groups due to repeated cases.  }
    \label{fig:case_selection}
\end{figure}

Case selection is a process guided by the objectives of the study. Since \textit{MammoClean} is designed to provide harmonized datasets for multi-view mammography analysis across tasks such as breast density classification, diagnostic labeling, and BI-RADS assessment, the initial data selection was performed with these goals in mind. For each dataset, the availability of complete data and the presence of potential inconsistencies between different views of the same laterality were carefully checked, and examinations with missing or inconsistent information were excluded. In CBIS-DDSM, data are organized image-based, and the inclusion and exclusion criteria applied at the examination level are illustrated in Figure \ref{fig:case_selection}. In TOMPEI-CMMD, imaging data are organized patient-based while metadata are provided examination-based; thus, no internal inconsistencies were observed, and only cases explicitly marked as excluded were removed from the study. In VinDr-Mammo, imaging data are organized patient-based and metadata image-based. As no critical inconsistencies were identified, all cases from this dataset were retained for analysis.

\subsubsection{Image Data Processing}

Processing of the imaging data began with normalizing the images to ensure a consistent dynamic bit range across all cases. Subsequently, each image was checked for laterality, and to maintain consistency, images with right laterality were horizontally flipped, resulting in a uniform orientation. Following this step, an additional procedure was applied to identify images with flipped intensity and correct them accordingly. Figure \ref{fig:MammoClean} illustrates the overall process with representative examples. Approximately 28\% of the images in the CBIS-DDSM dataset were identified with flipped laterality issues, and about 23\% of the images in the VinDr-Mammo dataset exhibited flipped intensity artifacts.

\subsubsection{Unified Data Storing}
To enhance reproducibility and facilitate future research, both the metadata and imaging data were unified and stored in a standardized format. Common features across the datasets were extracted and harmonized, while additional valuable features, such as the characteristics of abnormalities, even when not available in all studies, were preserved to maximize information utility. The newly unified metadata files contain the following information:

\begin{itemize}[noitemsep, topsep=0pt]
    \item \textit{ID:} Encoded patient identifier.
    \item \textit{Image ID:} Encoded image identifier, if applicable.
    \item \textit{Laterality:} Breast side depicted \{L, R\}.
    \item \textit{View:} Imaging view of the breast \{CC, MLO\}.
    \item \textit{Age:} Patient age.
    \item \textit{Breast Density:} Breast density category \{A, B, C, D\}.
    \item \textit{Diagnosis:} Biopsy-proven diagnostic label for benign and malignant cases \{Normal, Benign, Malignant\}.
    \item \textit{BI-RADS Assessment:} BI-RADS score \{1, 2, 3, 4, 5\}.
    \item \textit{Mass:} Presence of a mass abnormality \{0, 1\}.
    \item \textit{Mass Shape:} Shape of the mass, if applicable.   
    \item \textit{Mass Margin:} Margin of the mass, if applicable.
    \item \textit{Mass Density:} Density of the mass, if applicable.
    \item \textit{Calcification:} Presence of a calcification abnormality \{0, 1\}.
    \item \textit{Calcification Morphology:} Morphology of the calcification, if applicable.
    \item \textit{Calcification Distribution:} Distribution of the calcification, if applicable.
    \item \textit{Asymmetry:} Presence and type of asymmetry \{Asymmetry, Focal Asymmetry, Global Asymmetry\}.
    \item \textit{Architectural Distortion:} Presence of architectural distortion \{0, 1\}.
    \item \textit{Other Findings:} Additional abnormalities \{Skin Retraction, Skin Thickening, Nipple Retraction, etc.\}.
    \item \textit{Split:} Dataset partition \{train, test\}.
    \item \textit{Image File Folder (raw):} The main folder that contains image file in the raw dataset.
    \item \textit{Image File Path (processed):} Path to the image file in the processed dataset.
\end{itemize}

It is noteworthy that, due to the presence of patient overlap between the training and test sets in the CBIS-DDSM dataset, the predefined split was discarded, and a train/test split was only retained for the VinDr-Mammo dataset. Imaging data were organized at the patient level, where each folder contained the available images along with a text file containing metadata such as age, breast density, diagnosis, and BI-RADS assessment (laterality-based, if available). The imaging files were named using the \textit{Laterality\_View} format, and this structure enables consistent reading and evaluation of the data. 

\subsection{Bias Analysis}
To better analyze the datasets and identify potential biases that may influence the training and performance of deep learning models, we examined several aspects, including the distribution of data across the main classification tasks, their co-occurrence patterns, and the prevalence of abnormalities across different clinical decision-making levels. The reported numbers are based on the processed metadata, and some imaging files may be missing due to issues such as repository deletions or server-related problems.

\begin{figure}
    \centering
    \includegraphics[width=1\linewidth]{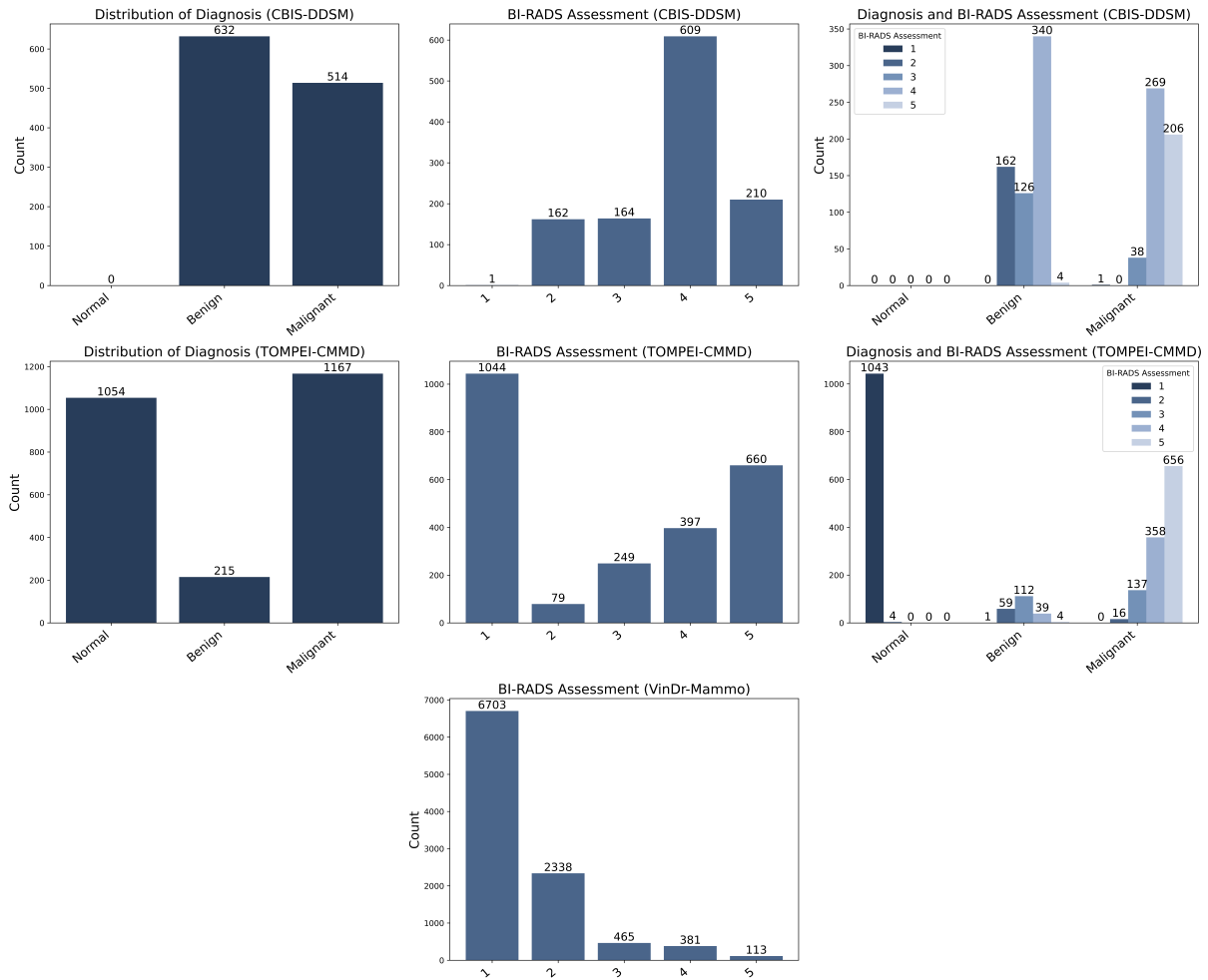}
    \caption{Data distribution for the diagnostic label and the BI-RADS score and their co-occurrence in different datasets after initial pre-processing.}
    \label{fig:Diagnosis-BIRADS}
\end{figure}

Figure \ref{fig:Diagnosis-BIRADS} presents the distribution of studies within the datasets for diagnostic labels, the BI-RADS scores assigned, and the distribution of these scores within each diagnostic category after the initial pre-processing. As illustrated, each dataset exhibits distinct biases toward specific conditions. In the CBIS-DDSM dataset, benign and malignant cases are relatively balanced; however, BI-RADS assessments are strongly skewed, with the majority of cases labeled as BI-RADS 4, and no representation for normal cases. In the TOMPEI-CMMD dataset, while malignant and normal cases are relatively balanced, benign cases are underrepresented. BI-RADS 1 was almost exclusively assigned to normal cases, with only one benign case reported, while BI-RADS 5 represents the second most common category. In contrast, the VinDr-Mammo dataset shows a substantial imbalance, with BI-RADS 1 and 2 accounting for more than 90\% of the studies. Such skewed distributions can introduce strong biases into model development, requiring careful handling; otherwise, models risk overfitting to dominant groups, and even evaluation metrics may produce misleading results.

For both TOMPEI-CMMD and CBIS-DDSM, diagnostic labels and BI-RADS scores were provided, and as expected, malignant cases were generally associated with higher BI-RADS scores compared to benign cases. Nevertheless, some discrepancies were observed: a small number of BI-RADS 5 cases had biopsy results confirming benign findings, while certain malignant cases were assigned unexpectedly low BI-RADS scores, such as 1 or 2. This highlights that although BI-RADS and biopsy results are correlated, inconsistencies exist, an issue often overlooked in studies that directly use BI-RADS categories to infer diagnostic labels.

\begin{figure}
    \centering
    \includegraphics[width=1\linewidth]{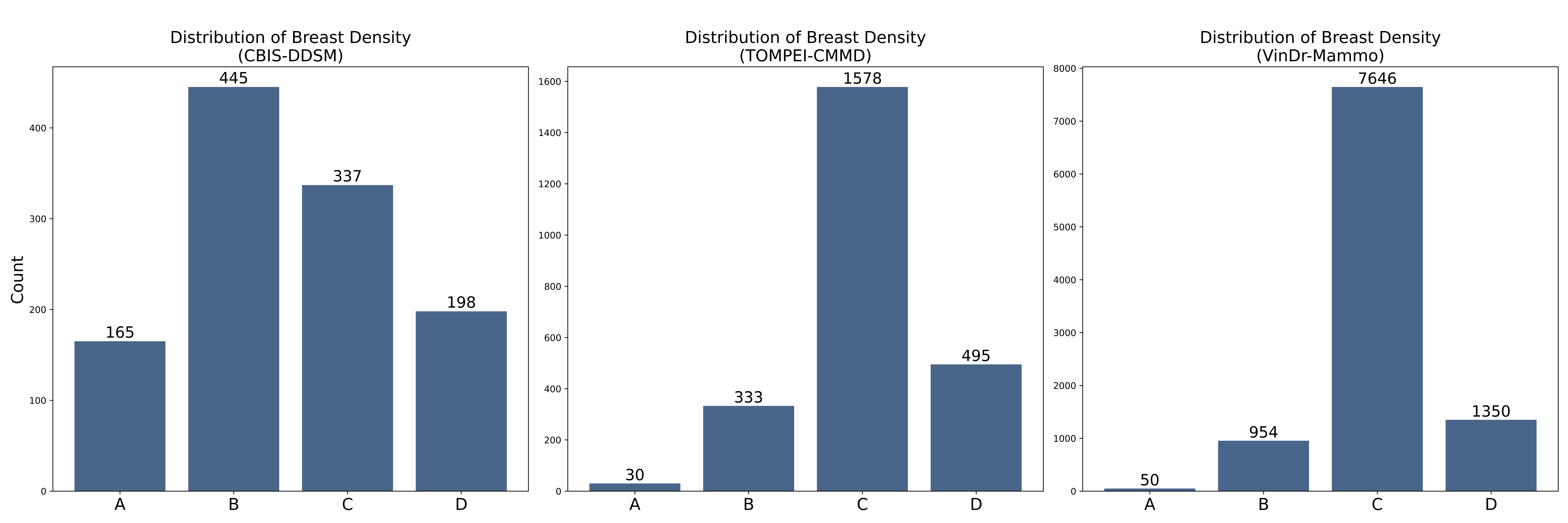}
    \caption{Breast density distributions across different datasets.}
    \label{fig:Breast-density-distribution}
\end{figure}

Figure \ref{fig:Breast-density-distribution} shows the distribution of breast density categories across the three datasets. In CBIS-DDSM, the majority of cases fall into category B, whereas in TOMPEI-CMMD and VinDr-Mammo, category C is dominant. This shift in distribution is expected, as CBIS-DDSM was collected in the United States, while the other two datasets were collected in China and Vietnam, reflecting known differences in breast density across ethnic groups. Such distributional shifts are clinically and technically important, as higher breast densities are associated with increased difficulty in abnormality detection, potentially affecting both radiological interpretation and model performance.

Table \ref{tab:abnormality-distribution} provides an overview of the distribution of abnormalities across different diagnostic and BI-RADS categories in the datasets. The results demonstrate substantial variation in abnormality distributions. For instance, while mass and calcification cases in CBIS-DDSM are almost balanced, in TOMPEI-CMMD the number of mass cases is approximately double that of calcifications, and in VinDr-Mammo the ratio is closer to three. Moreover, while most calcifications in CBIS-DDSM are associated with BI-RADS categories 2 and 4 and are predominantly benign, in TOMPEI-CMMD calcifications are primarily assigned to BI-RADS 5, with more than 86\% confirmed malignant. Asymmetry and architectural distortion cases are underrepresented across all datasets, consistent with their lower prevalence in clinical practice; however, compared to CBIS-DDSM and VinDr-Mammo, these categories are even more underrepresented in TOMPEI-CMMD. It should also be noted that VinDr-Mammo does not provide annotations for BI-RADS categories 1 and 2. Overall, these discrepancies underscore the importance of adopting careful sampling strategies to address class imbalance, a common challenge when training deep learning models with mammography data.

\begin{table}[ht]
\centering
\resizebox{1\textwidth}{!}{%
\begin{tabular}{llccccccccc}
\toprule
\textbf{Dataset} & \textbf{Abnormality} & \multicolumn{5}{c}{\textbf{BI-RADS score}} & \multicolumn{3}{c}{\textbf{Diagnosis}} & \textbf{Total} \\
\cmidrule(lr){3-7} \cmidrule(lr){8-10}
 &  & 1 & 2 & 3 & 4 & 5 & Normal & Benign & Malignant & \\
\midrule
\multirow{4}{*}{CBIS-DDSM} 
 & Mass & 1 & 24 & 121 & 255 & 136 & N/A & 266 & 271 &  537\\
 & Calcification & 0 & 143 & 43 & 357 & 77 & N/A & 371 & 249 & 620 \\
 & Asymmetry & 0 & 4 & 13 & 7 & 3 & N/A & 20 & 7 & 27 \\
 & Architectural Distortion & 0 & 4 & 0  & 35 & 16 & N/A & 13 & 42 & 55 \\
\midrule
\multirow{4}{*}{TOMPEI-CMMD} 
 & Mass & 1 & 6 & 162 & 308 & 533 & 1 & 122 & 887 & 1,010 \\
 & Calcification & 0 & 73 & 60 & 117 & 315 & 4 & 73 & 488 & 565 \\
 & Asymmetry & 1 & 0 & 25 & 6 & 0 & 0 & 18 & 14 & 32 \\
 & Architectural Distortion & 0 & 0 & 9 & 14 & 13 & 0 & 7 & 29 & 36 \\
 \midrule
 \multirow{4}{*}{VinDr-Mammo} 
 & Mass & N/A & N/A & 279 & 235 & 95 & N/A & N/A & N/A & 609 \\
 & Calcification & N/A & N/A & 32 & 118 & 79 & N/A & N/A & N/A & 229\\
 & Asymmetry & N/A & N/A & 157 & 72 & 14 & N/A & N/A & N/A & 243 \\
 & Architectural Distortion & N/A & N/A & 12 & 42 & 11 & N/A & N/A & N/A & 65 \\
\bottomrule
\end{tabular} }
\caption{ Distribution of abnormalities across diagnosis categories and BI-RADS assessments in different datasets. }
\label{tab:abnormality-distribution}
\end{table}

\section{Discussion}
In this paper, we discussed mammography images and their role in breast cancer screening and detection. We provided a comprehensive review of mammography analysis, beginning with the fundamentals of breast anatomy, breast density, mammographic findings, and the key factors influencing mammographic appearance, all of which can substantially affect the performance of deep models designed to automate image analysis and decision-making. We then evaluated the existing publicly available mammography datasets and observed that many of them lack the detailed findings and descriptors typically provided by radiologists during the diagnostic process. This limitation restricts the ability to evaluate model performance across subgroups defined by underlying conditions, making it difficult to identify model weaknesses and strengths and to mitigate them in order to develop fair and robust AI systems.

Another major observation is that datasets vary widely in the type and depth of information they contain, which limits their interoperability and complicates their joint use for model development. While such heterogeneity reflects the realities of different clinical environments, including variation in patient populations, imaging devices, and data collection protocols, it also poses critical challenges for reproducibility, comparability, and generalizability of AI models.

To address these challenges, we developed \textit{MammoClean}, a publicly available pipeline for pre-processing and harmonizing mammography datasets. \textit{MammoClean} was applied to three large-scale public datasets with annotations of underlying disease, CBIS-DDSM, TOMPEI-CMMD, and VinDr-Mammo, and is extendable to others. The framework standardizes both imaging and metadata across heterogeneous sources and incorporates quality-control steps such as laterality verification, intensity correction, and multi-view case selection. Our evaluation demonstrated that \textit{MammoClean} can resolve inconsistencies, unify dataset structures, and generate harmonized outputs suitable for AI applications.

By analyzing several factors in these datasets after harmonization, we showed that each dataset is biased toward specific conditions, often due to population-based shifts or case-selection practices. For instance, breast density distributions differ significantly across regions, underlining the importance of preserving demographic diversity in harmonized datasets. Moreover, discrepancies between BI-RADS assessments and biopsy-confirmed diagnoses emphasize the need for careful task design, particularly when BI-RADS categories are used directly as predictive targets. These differences, together with the imbalance in diagnostic categories and abnormality types, highlight the necessity of employing multiple datasets and adopting bias-aware training and evaluation strategies to avoid misleading results.

Although \textit{MammoClean} represents a step toward standardized mammography data preparation, limitations remain. While harmonization can reduce unnecessary variability, it cannot replace the need for larger and more diverse datasets, nor can it fully eliminate biases inherent in clinical practice and data collection. To further advance AI for breast cancer screening and diagnosis, we suggest several directions for future work.

\subsection{Subgroup-Specific Performance Assessment}
Most AI models are currently evaluated only on their overall outputs, without accounting for intra-category variations within each diagnostic group. However, considering these variations is critical for developing reliable and fair models suitable for clinical use. Subgroups can be defined not only by breast density, population characteristics, or age, but also by the underlying disease type and its associated features. Evaluating model performance across such subgroups is essential for identifying potential failure modes and for guiding the development of novel approaches, such as tailored pre-processing steps or adaptive learning strategies, that can mitigate these weaknesses.

\subsection{Toward Clinically Aligned AI Decision-Making}
In clinical workflow, the decision-making process follows a structured approach in which radiologists not only interpret mammographic images but also integrate additional information such as patient symptoms, age, and family history. In contrast, most AI models rely almost exclusively on imaging data. This limitation is partly due to the restricted availability of comprehensive metadata in existing datasets; however, even when such information is provided, it is rarely incorporated into model development. Another critical limitation of current AI-based approaches is their tendency to produce confident predictions without offering insight into the underlying reasoning or the specific abnormalities that informed the decision. This lack of transparency, combined with overconfident outputs and the black-box nature of deep learning models, hinders their trustworthiness and adoption in clinical practice. Ideally, AI systems should emulate the reasoning process of radiologists by identifying relevant findings and clearly explaining how these contributed to the final decision. Coupled with human-in-the-loop strategies, where radiologists can correct and guide model outputs, such approaches may not only improve model performance but also move AI systems closer to reliable integration into real-world clinical workflows.

\subsection{Directions for Future Standard Datasets}
Currently available datasets lack sufficient longitudinal data that are well organized in a time-ordered manner with step-specific decision labels and corresponding clinical recommendations. As a result, the development of models for risk assessment has remained limited compared to diagnostic models. Longitudinal datasets could enable dynamic modeling of breast changes over time and support the detection of subtle early-stage abnormalities, in contrast to the static approaches that rely on single-study images. Moreover, future dataset curation efforts should prioritize the inclusion of detailed lesion descriptors and richer metadata, allowing AI models to incorporate multi-source information more closely aligned with clinical workflows. Ensuring broader demographic representation is equally important to improve both the generalizability and the clinical applicability of AI systems.

Recent advances in medical AI are increasingly moving beyond unimodal image-based models toward multimodal frameworks that integrate diverse sources of patient information. In mammography, this trend involves combining imaging data with complementary modalities such as ultrasound, MRI, or digital breast tomosynthesis, as well as with clinical metadata including age, family history, genetic risk factors, and prior imaging studies~\cite{qian2025multimodal}. Multimodal integration can enhance diagnostic accuracy by capturing both morphological and contextual cues that are often considered by radiologists during routine decision-making, thereby narrowing the gap between AI systems and clinical reasoning.

A parallel development is the emergence of large-scale foundation models, pre-trained on massive heterogeneous datasets and adaptable to downstream clinical tasks~\cite{ghosh2024mammo}. These models exhibit promising transfer learning capabilities, enabling them to generalize across imaging devices, populations, and institutions. For breast cancer screening, foundation models that integrate multimodal inputs offer an avenue for more robust risk assessment, early detection of subtle changes across longitudinal exams, and alignment with clinical practice guidelines. Furthermore, the combination of foundation models with multimodal harmonized datasets, such as those enabled by \textit{MammoClean}, may accelerate the creation of scalable and interoperable AI systems.

Nevertheless, these advances also raise critical questions about computational cost, fairness, and interpretability. While multimodal models can reduce reliance on imaging alone and provide richer diagnostic insights, they may exacerbate disparities if auxiliary metadata are incomplete or systematically biased across subgroups. Similarly, foundation models demand rigorous evaluation to ensure that their broad adaptability does not compromise specificity or introduce spurious correlations. Addressing these challenges will require collaborative benchmarking, standardized evaluation protocols, and frameworks that integrate transparency and uncertainty quantification into model outputs.

\section{Conclusion}
In this work, we provided a comprehensive review of mammography analysis, beginning with the fundamentals of breast anatomy, breast density, and mammographic findings, followed by an evaluation of publicly available datasets and their challenges, and introduced \textit{MammoClean}, a reproducible and extensible pipeline for harmonizing mammography data. By standardizing imaging formats, metadata structures, and ensuring quality control through steps such as laterality verification, intensity correction, and multi-view consistency, \textit{MammoClean} resolves inconsistencies and unifies diverse datasets, as demonstrated on CBIS-DDSM, TOMPEI-CMMD, and VinDr-Mammo. Our comparative analysis highlighted critical insights, including regional differences in breast density distributions, discrepancies between BI-RADS assessments and biopsy-confirmed diagnoses, and imbalances in diagnostic categories that necessitate bias-aware training and evaluation strategies. While \textit{MammoClean} reduces unnecessary variability, the need for larger, more diverse datasets and careful task design remains essential. Overall, this study offers both a comprehensive overview of mammography analysis and a methodological contribution that lays the foundation for more consistent, reproducible, and clinically relevant AI applications, advancing the long-term goal of developing equitable and reliable tools for breast cancer screening and diagnosis.

\bibliographystyle{ieeetr}
\bibliography{refs.bib}
\end{document}